\documentclass[aps,prl,reprint,groupedaddress]{revtex4-1}

\usepackage{graphicx}
\usepackage{xcolor}

\begin{document}


\title{Fick's Second Law Transformed: One Path to Cloaking in Mass Diffusion}



\author{S. Guenneau}
\affiliation{Institut Fresnel, UMR CNRS 7249, Aix-Marseille Universit\'e, Campus de St J{\'e}r{\^o}me, 
13397 Marseille Cedex 20, France}
\author{T.M. Puvirajesinghe}
\affiliation{Institut Paoli-Calmettes, UMR INSERM 1068, \\
UMR CNRS 7258, Aix-Marseille Universit\'e, Marseille, France}

\def\sebApril{\bf}
\def\sebApril{}
\def\boris{\textcolor{blue}}
\def\seb{\textcolor{red}}

\date{\today}

\begin{abstract}
Here, we adapt the concept of transformational thermodynamics,
whereby the flux of temperature is controlled via 
anisotropic heterogeneous diffusivity, for the diffusion and
transport of mass concentration. The n-dimensional, time-dependent,
anisotropic heterogeneous Fick's equation is considered,
which is a parabolic partial differential equation also applicable to
heat diffusion, when convection occurs, for example in fluids.
This theory is illustrated with finite element
computations for a liposome particle surrounded by a
cylindrical multilayered cloak in a water-based environment,
and for a spherical multilayered cloak consisting
of layers of fluid with an isotropic homogeneous diffusivity,
deduced from an effective medium approach.
Initial potential applications could be sought in bio-engineering. 
\end{abstract}


\maketitle

\section{Introduction}
This communication  aims to target already existing systems which could enable the use of cloaking concepts in order to achieve control of three-dimensional processes using coated spheres consisting of concentric layers of homogeneous isotropic diffusivity. Various
applications already implicate the use of concentric bilayered vesicles, one example being liposomes used for drug delivery \cite{martins}. Liposomes are concentric bilayered vesicles in which an aqueous volume containing a water-soluble drug is enclosed by a membranous lipid bilayer composed of natural or synthetic phospholipids. One popular type of liposomes, known as the stealth liposomes \cite{stealth} are highly stable, long-circulating liposomes whereby polyethylene glycol (PEG) has been utilized as the polymeric steric stabilizer \cite{catalan}. Stealth and other liposomes use the concept of `invisibility' in order to hide and evade the immunosystem by coupling water-soluble polymers to the lipid heads. Therefore the polymer part of the molecule is dissolved in the aqueous environment, thus masking the liposomes from immune cells in the blood \cite{elbayoumi}. Other alternative applications to liposomes are nanoparticles based on solid lipids (SLN). These are composed of solid lipids stabilized with an emulsifying layer in an aqueous dispersion. This has benefits such as drug mobility.
The release of the drug-enriched core of SLN is based upon Fick's first law of diffusion \cite{muller}. 

\begin{figure}[h]
\resizebox{90mm}{!}{\includegraphics{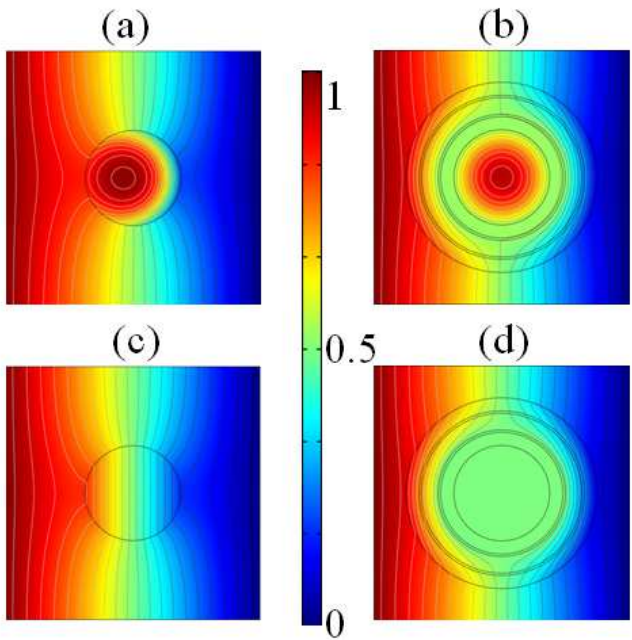}}
\caption{Two-dimensional simulation for diffusion of chemical species's concentration: Concentration is normalized to 1 mol.m$^{-3}$ on the left boundary with a flux boundary condition on right boundary with mass transfert coefficient of 5 m.s$^{-1}$, and symmetry boundary conditions on top and bottom; Two time point ($t=1\;10^{-6}$s,(a,c) and $t=1.5\;10^{-5}$s, (b,d)) simulations of mass diffusion in surrounding medium with diffusion constant of
$2.1 \; 10^{-9}$m.s$^{-2}$ (CO2-water) of a circular nano-size particle (nanobody) with a diameter of $1.5 \; 10^{-8}$m of diffusion
constant $1.9 \; 10^{-11}$m.s$^{-2}$(POPC-dehydrated).
(b,d): Application of a cloak, surrounding the nanobody, which is of inner radius $1.5 \; 10^{-8}$m and outer radius $3.0 \; 10^{-8}$m
and consists of 5 concentric layers. The first, third and fifth layers from the inside of the cloak outwards have a diffusivity of
$4.586 \; 10^{-10}$m.s$^{-2}$ (sucrose in water) and respective thickness $4.25 \; 10^{-9}$m, $5.25 \; 10^{-9}$m and  $4 \; 10^{-9}$m. The second layer and the fourth layers have diffusitivities of $11 \; 10^{-6}$m.s$^{-2}$ (Gaz ethanol-air) and $8.4 \; 10^{-9}$m.s$^{-2}$ (liquid CO2-Methanol) and identical thickness $7.5 \; 10^{-10}$m. Note that in panels (b) and (d) isovalues of concentration (black curves) are bent around the nanobody, whilst they remain aligned outside the cloak.
}
\label{fig1}
\end{figure}

Another similar idea to the one presented in this paper is the concept of optical transparency resulting from the use of preparative reagents for the subcellular localization of fluorescently labelled tissues and organisms \cite{hama}. High resolution imaging techniques such as Laser Scanning Microscopy (LSM) are able to provide high-resolution images of biological samples. However the resolution of imaging whole organisms such as embryos by fluorescently labelling certain components, becomes somewhat distorted due to the biological samples containing optically opaque regions. These opaque components are able to transmit, reflect, scatter as well as absorb light, which can lead to image distortions. Therefore certain commercially or non-commercially available reagents are available which are known as optical clearing reagents \cite{seisenberger,hama} and can render tissues and organisms transparent.

Here we suggest a novel application to the fast-growing research area of cloaking, whereby a better control of light can be achieved through transformational optics, following the pioneering theoretical works of Pendry et al. \cite{pendry06} and Leonhardt
\cite{leonhardt06}, to diffusion processes in biophysics. The aforementioned paper  \cite{pendry06} demonstrates the possibility of designing
a cloak that  renders any object inside it invisible to
electromagnetic radiation (using the covariant structure of Maxwell's equations),
while the latter \cite{leonhardt06} concentrates on the ray optics limit (using conformal mappings in the
complex plane for Schr\" odinger's equation). In both cases, the cloak consists of a meta-material
whose physical properties (permittivity and permeability) are spatially varying and
matrix valued.
This route to invisibility is reminiscent of the work of Greenleaf et al. in the context of electrical
impedance tomography \cite{greenleaf03}.
Interestingly, the isomorphism between the anisotropic
conductivity and thermostatic equations makes it
possible to control the pathway of heat flux
in a stationary setting, as observed by Fan et al.
\cite{huang} (see also \cite{ross} for analogous
cloaking in electrostatics) and experimentally validated
by Narayano and Sato  \cite{narayana}. However, time
plays an essential role in diffusion processes, and manipulation of
heat flux through anisotropic diffusivity requires greater care
in a transient regime \cite{sebcape,martin}. Interestingly, anisotropic
diffusion is a well-known technique in computer
vision \cite{giblin}
aiming to reduce image noise without removing significant
parts of the image content, typically edges,
lines or other details that may be important in the interpretation of the image
\cite{perona88}. Spatio-temporal differential equations of
a reaction-diffusion type also appear in organogenesis models
for the developments of limbs, lungs, kidneys and bones \cite{kondo}.


The mathematical model described in this report is based upon Fick's laws of diffusion,
derived by Adolf Fick in the year 1855 \cite{fick},
which describes diffusion processes governing various contexts (conduction of electricity, heat,
concentration of chemical species etc, and even gray scale image in computer vision).
Here, the diffusion coefficient is spatially varying (heterogeneous) and matrix valued (anisotropic).
We show numerically that control can be achieved of
three-dimensional processes (with a focus here on bio-engineering applications),
using coated spheres consisting of concentric layers of homogeneous isotropic diffusivity,
which mimick certain anisotropic heterogeneous diffusivity.
Previous studies have only shown the control of diffusion processes
with two-dimensional transformational thermodynamics \cite{huang,narayana,sebcape}.
Using a similar strategy, we extend earlier works to three-dimensional diffusion processes
and also discuss the issue of convection.

\section{Materials and Methods}

\subsection{Numerical simulations}
The time-dependent convection diffusion equation was implemented in the commercial finite element software COMSOL MULTIPHYSICS.
Two-dimensional computations such as the one shown in Fig. \ref{fig1}(b) were performed using a laptop with 3Gb of RAM memory, but only required 1Gb of RAM memory to achieve a complete convergence of the numerical solution to the parabolic problem (the complete convergence was checked by refining the mesh size and the time steps, in the limit of the 3Gb of available RAM memory). For instance, the mesh for the fully converged numerical simulation shown in Fig. \ref{fig1}(b) consisted of 16637 mesh points and 33142 triangular elements. This computation took less than two minutes using second order Lagrange equations and a time-dependent direct linear system solver (SPOOLES) with a symmetric matrix condition. Time step was $0.01s$ throughout the time interval $[0s,0.1s]$ with a relative tolerance $0.01$. However, three-dimensional computations shown in Fig. \ref{fig2}-\ref{fig3} were more computationally demanding than in the two-dimensional case. They were performed on a cluster of 16 computers with a total of 1Tb RAM memory. Nevertheless, fully converged numerical solutions required
only 36 Gb of RAM memory to properly run (convergence of the results was double-checked refining the mesh and time steps, what required up to 524Gb of RAM memory and 48 hours of computation). For the results shown in Fig. \ref{fig2}-\ref{fig3}, the mesh consisted of 152805 mesh points and 905294 tetrahedral elements for the 3D computations, and computations took approximately 12 hours (using up to 36Gb of RAM memory). More specifically, we used second order Lagrange elements and a congugate gradients time dependent linear system solver with an Algebraic multigrid (quality of multigrid hierarchy of 3) and time step of $0.01s$ throughout the time interval $[0s,0.1s]$, with a relative tolerance of $0.01$. 

\section{Results and discussion}
Recent work used a change of coordinates in the time-dependent heat equation \cite{fourier} to achieve a marked enhancement in the
control of heat fluxes in two-dimensional media described by an anisotropic heterogeneous conductivity \cite{sebcape}. However, it has been known since the work of Fick \cite{fick} that there is a deep analogy between diffusion and conduction of heat or electricity:
Because of Fick's work, diffusion can be described according to the same mathematical formalism as Fourier's law for heat conduction, or Ohm's law for electricity. We would like to use similar analogies between diffusion
of heat and concentration of chemical species to propose an original strategy towards
cloaking in bio-engineering/chemical engineering. A possible application is shown in Fig. \ref{fig1}, through the creation of different layers. This concept has already been used to a certain extent in multivesicular liposomes which consist of bilayers of phosholipids. Using the diffusivity values published for a certain lipid-conjugated drug ($1.9 \, 10^{-11}$m.$s^{-2}$ for dehydrated POPC) and coating these nanosized particles (30 nanometers) with different layers of sucrose  (diffusivity of $4.586 \, 10^{-10}$m.$s^{-2}$) and also thin layers containing substances of higher diffusivities, results in the initial particle becoming invisible in the chemical environment (diffusivity value of $2.1 \, 10^{-9}$m.$s^{-2}$), which is used to represent an environment such as blood whose main constituent plasma is essentially composed of water \cite{web1,web2}. This can be seen with the comparison of the distribution of concentration without the cloak, Fig. \ref{fig1}(a,c), and with the cloak,
Fig. \ref{fig1}(b,d). The consequences of the additional layers are more prominent at longer time points compared to shorter time points,
$t=1.5 \, 10^{-5}$s (Fig. \ref{fig1}, panels (c,d)) compared to $t=1 \, 10^{-6}$s (Fig. \ref{fig1}, panels (a,b)). Thus, the effect of
the different layers aids in maintaining high concentrations of a substance in the center of the liposome for longer
periods of time. This could have advantages in increasing drug stability for longer circulation times.	 

Potential fabrication of this alternative liposome would involve additional procedures to the classical liposome fabrication steps currently used. More specifically, chloroform or chloroform-methanol or mixing with the lipid and hydrophobic organic solvents could still be used for the formation of vesicles. The appropriate removal of solvents involving rotary evaporation for extended time periods could still be used. This could be followed by frozen storage before consequent steps. Secondly, replacing the classical hydration step by using an aqueous medium but with a higher concentration of sucrose compared to traditional sucrose concentrations (with mixtures of glycine or alanine, which share the same diffusion coefficients  and are also traditionally used for coating drugs), would allow the creation of sucrose layers during the formation of micelles.
Choosing appropriately sized vesicles would involve sonication techniques and analysis using techniques employed for vesicular structures such as dynamic light scattering (DLS) equipment and transmission electron microscopy (TEM). 

Layers containing the gaseous phase or an appropriate replacement of a high diffusivity value (in the range of $10^{-6}$ to $10^{-5}$m.$s^{-2}$) would be the most complicated steps in the fabrication process. In practice, it would be more feasible to replace the gaseous layers used in the numerical stimulations by media of similar diffusivity values, due to difficulties in initial fabrication and stability or maintenance of these layers. 

It should be noted that the examples of specific substances chosen for the simulations can easily be replaced by alternative appropriate substances with the same diffusion coefficient values. In addition, if potential cloaking applications are non-objectionable to the use of chloroform, it should be noted that this replacement has already been calculated to be a good substitute for the sucrose layers (layer one, three and five). This is shown in Supplementary Fig. 1, wherein there is improved cloaking as the isovalue curves for concentration outside the multilayered structure are nearly perfectly aligned (panels (b),(d)) in contradistinction to what can be observed in panels (a), (c) for a nanoparticle not surrounded by a cloak. Supplementary Fig. 2 shows that the maximum concentration
within the nanoparticle is always lower when it is surrounded by the cloak (and that its variation is dramatically reduced because of
the cloak). This is demonstrated by calculating the concentration at all points along a line passing through the
center of the nanoparticle without (Supplementary Fig.2, panels (a),(c)) and with (Supplementary Fig. 2, panels (b),(d)) a surrounding cloak. Note that this is achieved with simply 5 concentric layers, 3 of which have same diffusivity. A similar type of
profile for concentration can be observed in Supplementary Fig. 4, for a three-dimensional cloak with 20 layers
described in Supplementary Fig. 3.

\subsection{n-dimensional transformed convection-diffusion equation}
We consider the convection-diffusion equation which is a parabolic partial differential equation combining the diffusion equation and the advection equation. This equation describes physical phenomena where particles or energy (or other physical quantities) are transferred inside a physical system due to two processes: diffusion, which results in mixing and transport of chemical species without requiring bulk motion (it is a random walk of particles/molecules towards certain equilibrium state i.e. homogeneous distribution of chemical species inside a region), and convection, whereby collective movements of ensembles of molecules take place (usually in fluid) which in essence use bulk motion to move particles from one place to another place \cite{stanford}. In its simplest form (when the diffusion coefficient and the convection velocity are constant and there are no sources or sinks),
the convection-diffusion equation  in a domain $\Omega$ (with a chemical source outside) can be expressed as \cite{cussler}
\begin{equation}
\frac{\partial c}{\partial t}=\sum_{i,j}\frac{\partial}{\partial x_i}(\kappa_{ij}(x) \frac{\partial c}{\partial x_j})
- \sum_i \frac{\partial}{\partial x_i} v_i c \; ,
\label{heat1}
\end{equation}
where $c$ represents the mass concentration (in biochemistry) evolving with
time $t>0$, $\kappa$ is the chemical diffusion in units of $m^3.s^{-1}$, and $\underline{v}$ the velocity field.
We note that Fick's equation is written in a general form, where $x=(x_1,...,x_n)$ is a variable in
an n-dimensional space. Accordingly, sums stretch from $i,j=1,...n$ (here applications are sought
in 2d and 3d spaces, so $n=2$ or $3$).
It is customary to put matrix $\underline{\underline{\kappa}}$ in front of the spatial
derivatives when the medium is homogeneous.
However, here we
consider a heterogeneous (possibly anisotropic) medium,
hence the spatial derivatives of $\underline{\underline{\kappa}}$
might suffer some discontinuity (mathematically, partial derivatives
are taken in distributional sense \cite{jko95}, hence transmission conditions
ensuring continuity of the heat flux $\underline{\underline{\kappa}}\nabla c$
are encompassed in (\ref{heat1})). Physically, the diffusion flux
$-\underline{\underline{\kappa}}\nabla c$ measures the amount of substance that
will flow through a small volume during a short time interval
($mol.m^{-3}.s^{-1}$).


Upon a change of variable $x=(x_1,x_2,x_3)\to y=(y_1,y_2,y_3)$
described by a Jacobian matrix ${\bf J}$ such that $J_{ij}=\partial y_i/\partial x_j$,
(\ref{heat1}) takes the form:
\begin{equation}
\begin{array}{cc}
&\displaystyle{\frac{1}{\rm{det J_{ij}}}\frac{\partial c}{\partial t}}
=\displaystyle{\sum_{i,j}\frac{\partial}{\partial y_i}(\frac{1}{\rm{det J_{ij}}}J_{ij}
\kappa_{ij}(y) J_{ij}^T\frac{\partial c}{\partial y_j})} \nonumber \\
&\displaystyle{-\sum_{i,j}\frac{1}{\rm{det J_{ij}}}J_{ij}v_i c} \; ,
\end{array}
\label{heat2}
\end{equation}
where $\underline{\underline{\kappa'}}={\bf J}\underline{\underline{\kappa}}{\bf J}^{T}\hbox{det}({\bf J})^{-1}$ and $\underline{v'}=\hbox{det}({\bf J})^{-1}{\bf J}^{T}\underline{v}$ are the transformed diffusivity
and velocity, respectively.

\subsection{Jacobian matrix and transformed diffusivity for cloaking}
Let us now consider the following transform \cite{greenleaf03}
\begin{equation}
F(x)=\left ( 1+ \frac{1}{2} \mid x \mid \right) \frac{x}{\mid x\mid} \; ,
\label{ndgeo}
\end{equation}
where $\mid x\mid=\sqrt{x_1^2+x_2^2+..+x_n^2}$. This function
is smooth except at point $O=(0,..,0)$. It blows up the point $O$
to the hypersphere of radius $\mid x\mid=1$, while mapping the hypersphere of radius $\mid x \mid=2$
to itself. Moreover, $F(x)=x$ at the boundary $\mid x\mid=2$.

\noindent Defining the Jacobian matrix ${\bf J}$ as $J_{ij}=\partial F_i/\partial x_j$, we find:
\begin{equation}
{\bf J}=\left ( 1+ \frac{1}{2} \mid x \mid \right) {\bf I} -\frac{1}{\mid x\mid} \hat{x}\hat{x}^T\; ,
\end{equation} 
where ${\bf I}$ is the $n\times n$ identity matrix and $\hat{x}=x/\mid x\mid$. This
Jacobian is well defined everywhere except at $x=0$.

\noindent We note that ${\bf J}$ is symmetric, $\hat{x}$ is an eigenvector with eigenvalue $1/2$
and $\hat{x}^\perp$ is a $n-1$ dimensional eigenspace with eigenvalue $1/2+1/\mid x\mid$
in a space of dimension $n$.

\noindent The determinant of the Jacobian follows easily:
\begin{equation}
\rm{det}({\bf J})=\frac{1}{2}{\left ( 1+ \frac{1}{2} \mid x \mid \right)}^{n-1}
= \frac{(\mid x\mid+2)^{n-1}}{2^n\mid x\mid^{n-1}} \; .
\end{equation}

\noindent There are two cases of practical importance: $n=2$, for which
writing $r=\mid x\mid=\sqrt{x_1^2+x_2^2}=2(\mid y \mid-1)
=2(r'-1)$, one can see that the eigenvalues of the matrix
of transformed diffusivity $\underline{\underline{\kappa}'}$
behave like $r$ and $r^{-1}$
as $r\to 0$; $n=3$ for which
writing $r=\mid x\mid=\sqrt{x_1^2+x_2^2+x_3^2}=2(\mid y \mid-1)
=2(r'-1)$, one can see that the matrix of transformed
diffusivity $\underline{\underline{\kappa}'}$
has one eigenvalue which behaves like $r^2$ and two
like $r^{0}$ as $r\to 0$. Interestingly, for $n\geq 4$,
one eigenvalue of $\underline{\underline{\kappa}'}$
behaves like $r^{n-1}$ and the remaining $n-1$
behave like $r^{n-3}$. This shows that only the
case $n=2$ leads to a singular matrix of
diffusivity $\underline{\underline{\kappa}'}$
at the inner boundary of the cloak
(the circumferential eigenvalue becomes
infinite), a fact
already noticed in the context of cloaking
for electric impedance tomography \cite{greenleaf03,kohn}.
However, this matrix is always degenerate
at the inner boundary of the cloak,
irrespective of the space dimension.
Similarly, $\underline{v'}$ is a null vector on
the inner boundary of the cloak only
in space dimension $2$. This
analysis provides evidence that spherical
cloaks should be easier to construct than
circular cloaks.

\noindent The parameters of the bio-cloak
need to be further analysed
in polar and spherical coordinates in order to simplify
the cloak's design and the numerical
implementation.

\subsection{On the choice of reduced parameters for
a bio-cloak in polar and spherical coordinates}
We first note that if we multiply both sides of
(\ref{heat2}) by $\rm{det J_{ij}}$ and let
$\rm{det J_{ij}}$ inside the partial space
derivatives, we retrieve the usual form
of the convection-diffusion equation,
albeit with anisotropic coefficients.
We realise this is not legitimate in
general as by doing so we add an extra
term
$\sum_{i,j}J_{ij}\kappa_{ij}(y) J_{ij}^T\frac{\partial}{\partial y_i}(\rm{det J_{ij}}
\frac{\partial c}{\partial y_j})$
in (\ref{heat2}), but we numerically checked that this term
can be sufficiently small that it does not significantly affect
the solution of the original
transformed equation (\ref{heat2}).
Physically, this manipulation results in preserving the direction
of the diffusion flux $-\underline{\underline{\kappa'}}\nabla c$
 (since $\rm{det J_{ij}}$ is a scalar).
However, it affects its continuity
(since $\rm{det J_{ij}}$ is heterogeneous).
Such a manipulation is known in the transformational optics
community to lead to transformed equation with
reduced parameters \cite{schurig06}.
We now observe that
from the function $F(r)=R_1+r(R_2-R_1)/R_2$ counterpart
of (\ref{ndgeo}), wherein $R_1=1$ and $R_2=2$,
in polar $(r,\theta)$ (resp. spherical
$(r,\theta,\phi)$) coordinates,
which blows up a point $O$ to the disc
(resp. the sphere) of radius $R_1$
and maps the disc (resp. the sphere) of radius
$R_2$ to itself
in polar $(r',\theta')$ (resp. spherical $(r',\theta',\phi')$)
coordinates,
the transformed diffusivity
can be expressed for a cylindrical cloak as:

\begin{equation}
\begin{array}{lll}
\kappa'_{r'} &=\displaystyle{{\left(\frac{R_2}{R_2-R_1}\right)}^2{\left(\frac{r'-R_1}{r'}\right)}^2} \; ,
& \kappa'_{\theta'}=\displaystyle{{\left(\frac{R_2}{R_2-R_1}\right)}^2} \; ,
\end{array}
\label{rhort1a}
\end{equation}

\noindent and for a spherical cloak as

\begin{equation}
\begin{array}{lll}
\kappa'_{r'} &=\displaystyle{{\left(\frac{R_2}{R_2-R_1}\right)}^4{{\left(\frac{r'-R_1}{r'}\right)}^4}} \; , \nonumber \\
& \kappa'_{\theta'}=\kappa'_{\phi'}=\displaystyle{{\left(\frac{R_2}{R_2-R_1}\right)}^4{{\left(\frac{r'-R_1}{r'}\right)}^2}} \; ,
\end{array}
\label{rhort1b}
\end{equation}
where $R_1$ and $R_2$ are the interior and the exterior radii of the cloak.
One note that when $r'$ tends to $R_1$, $\kappa'_{r'}$ goes to
zero and $\kappa'_{\theta'}$ remains constant in (\ref{rhort1a}),
whereas $\kappa'_{r'}$, $\kappa'_{\theta'}$ and $\kappa'_{\phi'}$
all go to zero in (\ref{rhort1b}). This means that thanks to the reduced
coefficients the matrix of transformed
diffusivity $\underline{\underline{\kappa}'}$
now has one eigenvalue which behaves like $r^2$ and one like $r^{0}$
as $r\to 0$ in the cylindrical case i.e. we no longer have
an eigenvalue which blows up on the inner boundary of the cloak.
Likewise, we now have one eigenvalue which behaves like
$r^4$ and two like $r^{2}$, instead of one behaving like
$r^2$ and two like $r^{0}$ when $r$ tends to zero,
if we use reduced parameters in the spherical case.
Thus, reduced parameters are an obvious
choice in the cylindrical case,
since $\kappa'_{\theta'}$ is a constant in  (\ref{rhort1a}),
and were implemented for thermal cloaks in \cite{sebcape,martin}.
However, in spherical case, $\kappa'_{\theta'}=\kappa'_{\phi'}$
are no longer constant in (\ref{rhort1b}). Nevertheless, we
need use such reduced coefficients to get rid of the
coefficient sitting in front of the time derivative in (\ref{heat2}).

\noindent These heterogeneous anisotropic (reduced or not)
parameters can be approximated
by piecewise constant isotropic coefficients, making
use of an effective medium approach,
as detailed in the supplementary material, which justifies the
implementation of multilayered cylindrical and spherical cloaks
with concentric isotropic homogeneous thin layers.
The choice of reduced parameters led in \cite{sebcape}
to a multilayered thermal cloak with piecewise constant
and high contrast diffusivity,
and we refer to values of diffusivity and computations therein
for the cloaking effect for a concentration of chemical species in
2D with 20 layers (there is a one-to-one correspondence between Fourier's heat equation
and Fick's equation, the unknown being either the temperature
or the concentration). However, Fig. \ref{fig1} of the
present paper clearly shows cloaking can be achieved
with simply 5 layers with moderate contrast in
diffusivity. Moreover, in the spherical case, the choice
of reduced parameters leads to a different set of parameters
(see supplemental material) for the multilayered bio-cloak,
which we now study numerically. 


\subsection{Numerical illustation}
\begin{figure}[h]
\resizebox{75mm}{!}{\includegraphics{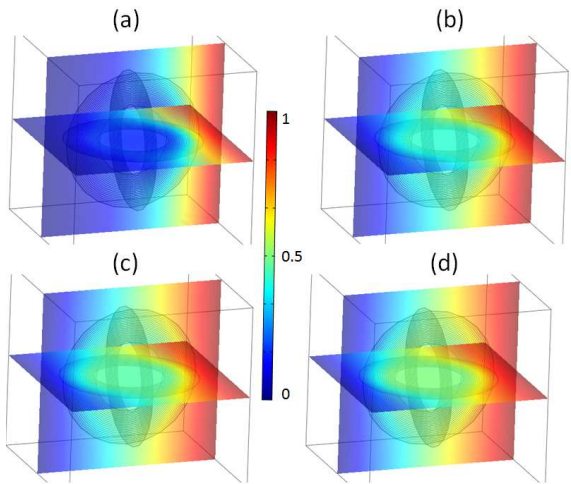}}
\caption{3D plot of concentration [mol/m$^3$]: (a) t=0.005s; (b) t=0.01s; (c) t=0.015s; (d) t=0.025s.
It has been checked that 3D plots are as in (d) for $t>0.025$s (steady state).
Spherical cloak of inner radius $1.5 \; 10^{-6}$m and outer radius $3.0 \; 10^{-6}$m consists of
20 concentric layers with diffusity ranging from $2.5 \; 10^{-6}$m$^2$.s$^{-1}$ to $1.7 \; 10^{-2}$m$^2$.s$^{-1}$.
The core and outer medium have same diffusivity $1.5 \; 10^{-5}$m$^2$.s$^{-1}$.}
\label{fig2}
\end{figure}

\begin{figure}[h]
\resizebox{75mm}{!}{\includegraphics{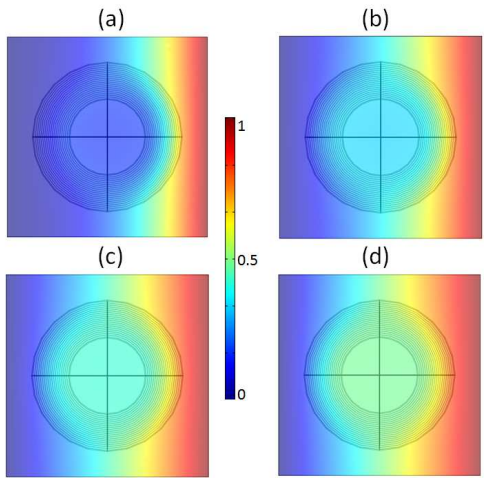}}
\caption{2D plot of concentration [mol/m$^3$] corresponding to a slice of 3D plot in Fig. \ref {fig2} in the horizontal plane passing through the center of the cloak: (a) t=0.005s; (b) t=0.01s; (c) t=0.015s; (d) t=0.025s.
It has been checked that 2D plots are as in (d) for time steps $t>0.025s$ (steady state).}
\label{fig3}
\end{figure}

For illustrative purposes, we focus here on a spherical cloak
consisting of $20$ concentric layers with diffusitivity ranging
from $2.5 \; 10^{-6}$m$^{2}$.s$^{-1}$ to $1.7 \; 10^{-2}$m$^2$.s$^{-1}$,
further details can be found in the supplemental material.
In order to emphasize the power of the approach,
we now consider a cloak substantially larger than that in
Fig. \ref{fig1}. Indeed, as observed in \cite{martin} in
the context of the transformed heat equation,
if we rescale the coordinate system as
$x\to sx$ and the time variable
$t\to s^2t$ with a dimensionless factor
$s$, the transformed Fick's equation
(\ref{heat2}) remains unchanged,
provided we assume velocity
is ruled out.
More precisely, the cloak
of Fig. \ref{fig1}
has been scaled up by a factor $s=10^2$ i.e.
its inner radius $1.5 \; 10^{-6}$m and
its outer radius $3.0 \; 10^{-6}$m
in Fig. \ref{fig2}.
Accordingly, time should be
scaled up by a factor $s^2=10^4$.
We show the distribution of
concentration for a
selectin of time points ranging
from $t=5 \, 10^{-3}$s to
$t=2.5 \, 10^{-2}$s in
Fig. \ref{fig2}.
Moreover, in Fig. \ref{fig2},
the cloak is in presence of a chemical specy with concentration
normalised to $1$ mol.m$^{-3}$ for simplicity (taking any other concentration $C$ will
simply lead to a color scale in Fig. \ref{fig2} and \ref{fig3} ranging from $0$ to $C$
mol.m$^{-3}$), 
which is set on the right-hand side of the computational domain
(a cube of sidelength $8.0 \; 10^{-6}$m). On the opposite (left-hand) side, we set the usual
flux condition $(\kappa\nabla c)\cdot {\bf n}=N_0+k_c (c_b-c)$ (with $c$ the,
as yet unknown, solution to (\ref{heat1}) and ${\bf n}$ the unit outward normal
to each side of the cube), where $\kappa$ varies within the range given
in the caption of Fig. \ref{fig2} inside the layers of the cloak (see supplemental material
for more details), and $\kappa=1$ m$^{2}$.s$^{-1}$ in the inner core
and outside the spherical cloak and $k_c$ is
the mass transfer coefficient, $N_0$ is the inward flux and $c_b$ is the bulk
concentration of chemical species in the cubic domain. Here,
we consider $k_c=5$ mol.s$^{-1}$, $N_0=0$ mol.m$^{-2}$.s$^{-1}$
and $c_b=0$ mol.m$^{-3}$. Finally, we set insulation (or equivalent symmetry)
conditions $(\kappa\nabla c)\cdot {\bf n}=0$ on the four remaining sides of the cube.
More details on implementation of such a diffusion model
in finite elements may be found in \cite{morton}.
This phenomenological model of a bio-cloak exhibits the following features:
The concentration of chemical species nearly vanishes inside the inner sphere of the
bio-cloak at time $t=0.005$s, see Fig. \ref{fig2}(a). In the optical setting, such
a sphere is called invisibility region, as no scattering obstacle placed inside
this region can be detected \cite{leonhardt06,pendry06,greenleaf03}.
In a bio-physical setting, this zone acts as a protection from any potential
chemical attack. However, the concentration of chemical species increases
steadily over time until it reaches half the value of the concentration which is set on
one side of the cubic computational domain, see Fig. \ref{fig2}(d).
We note that the concentration is
always uniform inside this inner sphere at any time point, see  Fig. \ref{fig2}(a)-(d).
Such a bio-cloak therefore offers some kind of protection from attack of chemical
species since the concentration is uniform in its inner sphere at any time point
and the concentration therein at any time point is always smaller than it
would be without a cloak (see supplemental material). Moreover, it
never exceeds a value, which in our configuration
is half the applied concentration (it can be seen
that this is due to the fact that the cloak has 
its origin in the center of the cubic domain).
   
\section{Conclusion}
In conclusion, in this brief communication we introduce a coordinate transformation
approach to control diffusion processes via anisotropy with an
emphasis on concentration of chemical species for potential applications in
bio-physics or bio-engineering. Not only does the form of the transformed
convection-diffusion equation involve an anisotropic
heterogeneous diffusivity, but it also requires a spatially varying coefficient
in front of the time derivative, as well as an anisotropic heterogeneous
velocity field. In order to be able to design a structured cloak for such
diffusion processes, we have simplified this transformed equation by
introducing so-called reduced coefficients, which preserve the
direction of the diffusion flux (but create an impedance mismatch
between the cloak and surrounding medium) and by further
assuming a small velocity in a homogenization approach. 
Our theoretical results are exemplified with numerical
simulations in two- and three-dimensional settings.
It should be emphasized that the issue of
convection,
which is of particular importance for
diffusion in fluids, such as in living organisms,
will require further comprehensive studies.

\begin{acknowledgments}
S.G. is thankful for a European funding (ERC starting grant ANAMORPHISM).
T.M.P. was funded by Fondation de La Recherche M\'edicale. 
\end{acknowledgments}


\begin{thebibliography}{99}
\bibitem{martins}
Martins, S., Sarmento, B., Ferreira, D. C., Souto, E. B. 2007
Lipid-based colloidal carriers for peptide and protein delivery--liposomes versus lipid nanoparticles.
{\it Int J Nanomedicine.} 2(4): 595-607.
\bibitem{stealth}
The Danish National Research Foundation for Biomembrane Physics (MEMPHYS)
at the University of Southern Denmark, Odense. Bioimaging, Science in Your Eyes. 2008 http://www.scienceinyoureyes.com/index.php?id=80.
\bibitem{catalan}
Barrajon-Catalan, E., Menendez-Gutierrez, M.P., Falco, A., Saceda, M., Catania, A. and Micol, V. 2011.
Immunoliposomes: A Multipurpose Strategy in Breast Cancer Targeted Therapy  in
{\it Breast Cancer - Current and Alternative Therapeutic Modalities} (Ed. E. Gunduz and M. Gunduz), ISBN 978-953-307-776-5, InTech.  
\bibitem{elbayoumi}
Elbayoumi, T.A., Torchilin, V.P. 2006 Enhanced accumulation of long-circulating
liposomes modified with the nucleosome-specific monoclonal antibody 2C5 in
various tumours in mice: gamma-imaging studies.  {\it Eur. J. Nucl. Med. Mol. Imaging} 33(10):1196-1205.
\bibitem{muller}
Muller, R.H., Dingler, A., Weyhers, H. Muller, R.H., zur Muhlen, A. 1997
Feste lipid nanopartikel (SLN). In {\it Pharmazeutische Technologie: Moderne
Arzneiformen}, (Ed. R.H. Muller, G.E. Hildebrand), pp. 265-272. Stuttgart: Wissenschaftliche Verlagsgesellschaft
\bibitem{hama}
Hama H., Kurokawa H., Kawano H., Ando R., Shimogori T.,
Noda, H., Fukami, K., Sakaue-Sawano A., Miyawaki A.
2011 Scale: a chemical approach for fluorescence imaging and
reconstruction of transparent mouse brain.
{\it Nat. Neurosci.} 14(11):1481-1488
\bibitem{seisenberger} 
Seisenberger, G., Ried M.U., Endress, T., Buning H., Hallek, M., Brauchle, C. 2001
Real-Time Single-Molecule Imaging of the Infection Pathway of an
Adeno-Associated Virus,
{\it Science} 294 (5548): 1929-1932
\bibitem{pendry06}
Pendry, J.B., Schurig, D. and Smith, D.R. 2006
Controlling Electromagnetic Fields,
{\it Science} 312, 1780
\bibitem{leonhardt06}
Leonhardt, U. 2006
Optical Conformal Mapping,
{\it Science} 312, 1777
\bibitem{greenleaf03}
Greenleaf, A., Lassas, M. and Uhlmann, G. 2003
On nonuniqueness for Calderon's inverse problem,
{\it Math. Res. Lett.} 10, 685-693
\bibitem{huang}
Fan, C.Z., Gao, Y. and Huang, J.P. 2008
Shaped graded materials with an apparent negative thermal conductivity
{\it Appl. Phys. Lett.} 92, 251907.
\bibitem{ross}
Nicorovici, N.A.P., Milton, G.W., McPhedran, R.C. and Botten, L.C. 2007
Quasistatic cloaking of two-dimensional polarizable
discrete systems by anomalous resonance,
{\it Opt. Express} 15, 6314-6323.
\bibitem{narayana}
Narayana, S. and Sato, Y. 2012
Heat flux manipulation with engineered thermal materials,
{\it Phys. Rev. Lett.} 108, 214303.
\bibitem{sebcape}
Guenneau, S., Amra, C. and Veynante, D. 2012,
Transformation thermodynamics : cloak and concentrator for heat,
{\it Opt. Express} 20(7): 8207-8218.
\bibitem{martin}
Schittny, R., Kadic, M., Guenneau, S. and Wegener, M. 2012
Experiments on transformation thermodynamics: Molding the flow of heat
(arXiv:1210.2810)
\bibitem{giblin}
Hallinan, P.L., Gordon, G.G., Yuille, A.L., Giblin, P. and Mumford, D. 1999
{\it Two and three dimensional patterns of the face},
Natick, Massachusetts: A.K.Peters.
\bibitem{perona88}
Perona, P. and Malik, J. 1990,
Scale-space and edge detection using anisotropic diffusion,
{\it IEEE Transactions on Pattern Analysis and Machine Intelligence}
12(7): 629-639.
\bibitem{kondo}
Kondo, S. and Miur, T. 2010
Reaction-diffusion
model as a framework for understanding
biological pattern formation.
{\it Science} 329: 1616-1620.
\bibitem{fick}
A. Fick 1855
On Liquid Diffusion.
{\it Philosophical Magazine} 4(10): 30-39.
\bibitem{fourier}
Fourier, J. 1822
{\it Th\'eorie analytique de la Chaleur},
Paris: Firmin-Didot p\`ere et fils.
\bibitem{web1}
Martinez, Isidoro. Mass Diffusivity Data. 1995-2013. http://webserver.dmt.upm.es/~isidoro/
\bibitem{web2}
Daykin, C.A. , Savage, A. K., Wulfert, F. 2013 Univeristy of Nottingham and Engineering and Physical Sciences Research Council (EPSCR).
i-metabolomics: spectra and diffusion coefficients of common metabolites found in blood plasma. http://www.nottinghamcourses.com/pharmacy/documents/i-metabolomics-spectra.pdf
\bibitem{stanford}
Stanford, A.L. 1975
{\it Foundations of biophysics},
 New-York, London: Academic Press
\bibitem{cussler}
Cussler, E.L. 1997
{\it Diffusion Mass Transfer in Fluid Systems},
New-York: Cambridge University Press.
\bibitem{jko95}
Jikov, V.V., Kozlov, S.M. and Oleinik, O.A., 1994
{\it Homogenization of Differential Operators and Integral Functionals},
New-York: Springer-Verlag
\bibitem{kohn}
Kohn, R.V., Shen, H., Vogelius, M.S. and Weinstein, M.I. 2008
Cloaking via change of variables in electric impedance
tomography, {\it Inverse Problems} 24: 015016.
\bibitem{schurig06}
Schurig, D., Mock, J.J., Justice, B.J., Cummer, S.A., Pendry, J.B.,
Starr, A.F. and Smith, D.R. 2006
Metamaterial electromagnetic
cloak at microwave frequencies.
{\it Science} 314: 977-80.
\bibitem{morton}
Morton, K.W. 1996
{\it Numerical Solution of Convection-Diffusion problems},
London: Chapman and Hall
\bibitem{wolfe}
Wolfe, C.A, James, P. A, Mackie, A, R., Ladha, S., Jones, R. 1998
Regionalized Lipid Diffusion in the Plasma Membrane of Mammalian Spermatozoa. {\it Biology of reproduction} 59, 1506-1514.
\bibitem{}
Gaede, H.C and Gawrisch, K. 2003
Lateral Diffusion Rates of Lipid, Water, and a Hydrophobic Drug in a Multilamellar Liposome.
{\it Biophys J.} 85 (3): 1734-1740. 
\bibitem{kwong}
Kwong P.D., Wyatt R., Robinson J., Sweet R.W., Sodroski J., Hendrickson, W.A. 1998
Structure of an HIV gp120 envelope glycoprotein in complex with the CD4
receptor and a neutralizing human antibody. {\it Nature} 393 (6686): 648-659 
\end{thebibliography}

\appendix

\end{document}